\documentclass[twocolumn,showpacs]{revtex4}

\usepackage{graphicx}
\usepackage{dcolumn}
\usepackage{amsmath}

\begin{document}

\title[Short Title]{Kondo Effect in Multiple-Dot Systems}

\author{Rui Sakano}
\author{Norio Kawakami}
\affiliation{
Department of Applied Physics,
Osaka University,  Suita, Osaka 565-0871, Japan\\
}
\date{\today}
\pacs{
68.65.Hb, 
71.27.+a, 
72.15.Qm, 
75.20.Hr 
}

\begin{abstract}
We study the Kondo effect in multiple-dot systems for which
the inter- as well as intra-dot Coulomb repulsions are strong,
and the inter-dot tunneling is small.
The application of the Ward-Takahashi
identity to the inter-dot dynamical susceptibility
enables us to analytically calculate the conductance for a
double-dot system by using the Bethe-ansatz exact solution of the
SU(4) impurity Anderson model. It is clarified
how the inter-dot Kondo effect enhances or suppresses
the conductance under the control of
the gate voltage and the magnetic field. We
then extend our analysis to multiple-dot systems including
more than two dots, and discuss their characteristic transport
properties by taking a triple-dot system as an example.
\end{abstract}

\maketitle

\section{Introduction}

Recent advances in semiconductor processing have made it possible
to fabricate various nanoscale materials with tunable
quantum parameters, revealing various aspects
of quantum mechanics.
Quantum dot
\cite{Tarucha,Reimann}
is one of the interesting nanoscale materials.
In particular, a lot of works on the Kondo effect in single
quantum dot systems have been done both theoretically and
experimentally
\cite{TK,Glazman,Kawabata,Izumida,Hofstetter,Eto,Oguri,DGG,Sasaki,Cho}.
More recently, double-dot systems or systems with more than
two dots have been
investigated
\cite{Nagaraja,WGW,Saraga}.
In this connection, the  Kondo effect in
double-dot systems have been studied intensively
\cite{Aono,Georges,Jeong,Taka,Tanaka,Martin}.

Most of the previous studies on multiple-dot systems have treated the
intra-dot Coulomb repulsion but have ignored the Coulomb repulsion
between quantum dots (inter-dot Coulomb repulsion). We especially focus
on the effect of the inter-dot Coulomb repulsion here
\cite{Wilhelm,Sun,Borda}
and study how such electron correlations affect transport properties.
Recently, Borda \textit{et al.} have investigated properties of the Kondo
effect in such double-dot systems with a magnetic field by
the numerical renormalization group method
\cite{Borda}, which may explain the  Kondo effect observed experimentally
 by Wilhelm \textit{et al.} in a double-dot system\cite{Wilhelm}.
A remarkable point in the above double-dot systems with
inter-dot Coulomb
repulsion is that enhanced charge fluctuations between the quantum dots
induce the ``inter-dot Kondo effect", which
plays an important role to determine transport properties of
the systems.  Since the inter-dot Kondo effect is caused
not by spin fluctuations but by charge fluctuations
between two dots, its influence appears
significantly when the dots are connected in series.
In particular, by changing the gate voltage or
the magnetic field, one can control the conductance
via the inter-dot Kondo effect.

In this paper, we investigate transport properties of
the double-dot systems with strong intra- and
inter-dot Coulomb repulsions mentioned above. We exploit a novel
method to treat the Kondo effect at absolute zero: the
application of the Ward-Takahashi identity enables us to
use the Bethe-ansatz exact solution of the
SU(4) impurity Anderson model to our double-dot system.
Our calculation clearly shows that the inter-dot Kondo
effect plays an important role on transport, which can be
controlled by the gate voltage and the magnetic field.
Our method is also applicable to multiple-dot systems including
more than two dots.  We explore the
Kondo effect in such dot systems by taking a triple-dot
system as an example.

This paper is organized as follows.
In section \ref{sc:model}, we introduce the model Hamiltonian and
outline the method to treat our double-dot system:
how  the Bethe-ansatz exact solution can be used to
compute the conductance at absolute zero. In section
\ref{sc:doubl-dotresult} we
discuss the results for the conductance
with particular emphasis on the gate-voltage
control and the magnetic-field control.
In section \ref{sc:maltiple} we extend our method to a
triple-dot system, and discuss its transport properties
on the basis of the exact solution. We also
mention
how we can treat generalized multiple-dot systems including
more dots. A brief summary is given in
section \ref{summary}.

\section{Model and Method}\label{sc:model}

\begin{figure}[bp]
\includegraphics[width=5.0cm]{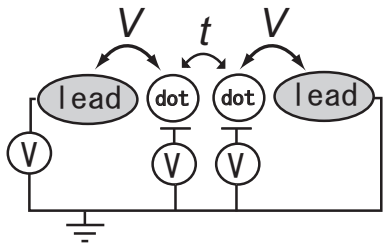}
\caption{Schematics of our double-dot system: two dots are connected via tunneling $t$, and each dot is connected to a lead via $V$.}
\label{fig:model}
\end{figure}

We describe our model and method by taking
a double-dot system connected in series, which was
proposed by Borda \textit{et al.} \cite{Borda}.
The setup is schematically shown in FIG. \ref{fig:model},
where not only the ordinary Coulomb repulsion $U$, which works inside each
dot, but also $U'$ between the dots are introduced. We assume that
the inter-dot tunneling $t$ is small and the gate voltages are
such that the lowest-lying charged states are restricted
to the configurations of singly-occupied states, $(n_R,n_L)=(1,0)$
and $(0,1)$, where $n_{R(L)}$ is the number of extra electrons on the
right (left) dot. This situation is realized in the condition,
\begin{equation}
|E(1,0)|, |E(0,1)| \ll U, U'
\end{equation}
where $E(n_R,n_L)$ is the energy level in the dots
measured from the common
chemical potential of the two leads. The
states $(1,0)$ and $(0,1)$ have a spin $S=1/2$, associated with
the extra electron on the double dots. Then at energies below the
charging energy of the double dots, dynamics of the double dots
is restricted to the subspace with the
4 possible configurations of $\{ S_z=\pm 1/2 ; \ n_R-n_L = \pm 1 \}$
in addition to the unoccupied state of $n_R=n_L=0$.

The above double-dot system, in which
both of intra- and inter-dot Coulomb repulsions are
sufficiently strong,  may be modeled by
the highly correlated degenerate Anderson Hamiltonian
$H_A$ ($U,U'\to\infty$) supplemented by  a inter-dot tunneling
term $H_T$,
\begin{eqnarray}
H_A&=& \sum_{\sigma,\tau} \int dx\, c_{\sigma \tau}^{\dagger}(x)
  \frac{1}{i}\frac{\partial}{\partial x}c_{\sigma \tau}(x)
    +  \sum_{\sigma,\tau}\varepsilon_{d\tau}^{\sigma}
    |\sigma\tau \rangle \langle \sigma \tau | \nonumber \\
&+&     V\sum_{\sigma\tau} \int dx\, \delta(x)
   [ | \sigma\tau \rangle \langle 0|c_{\sigma \tau}(x) + h.c.] , \\
\label{anderson}
H_T&= t & \sum_{\sigma,\tau\ne\tau'} | \sigma\tau \rangle \langle
\sigma \tau' |
\label{tunnel}
\end{eqnarray}
where $c_{\sigma \tau}^{\dagger}(x)(c_{\sigma \tau}(x))$ creates
(annihilates) a conduction electron at a position
$x$ with spin $\sigma$$(=\pm1/2)$ and
``orbital index" $\tau$. Here we have represented
conduction electrons in the leads in the low-energy continuum limit by
assuming that its density of states is constant, $1/2\pi$. Also we have
introduced the orbital index  $\tau=1/2$ (-1/2) to specify
an electron occupying the left (right) lead, which is
also used to label the left (right) dot.
A state  $| \sigma\tau \rangle $  in the double dots
located at $x=0$ denotes a singly occupied
state and  $|0\rangle$ denotes an unoccupied state.

We will discuss transport properties of the system under the
gate-voltage control or the magnetic-field control.
It is thus convenient to
write down each energy level $\varepsilon_{d\tau}^{\sigma}$  as,
\begin{eqnarray}
\varepsilon_{d\tau}^{\sigma} = \varepsilon + \delta E \cdot \tau + E_Z \cdot \sigma,
\end{eqnarray}
where $\delta E$ ($E_Z$) is the energy difference between
the two dots (Zeeman
energy). Note that the system possesses SU(4) symmetry
with respect to spin and orbital degrees of freedom
at $\delta E =E_Z =0$.

We note here that
the Bethe-ansatz exact solution can be obtained for
the above four-component Anderson Hamiltonian $H_A$ \cite{schkawa},
which is referred to as the SU(4) Anderson model henceforth.
However, this method allows us
to calculate only static quantities, so that we cannot
apply the exact solution to transport quantities straightforwardly.
In the following, we outline
how we can overcome this difficulty to calculate the conductance.

Let us begin with the expression for the conductance
in the above double-dot system connected in series, which is
obtained in the second order in the tunneling Hamiltonian $H_T$
between two dots,
\begin{equation}
G=-\frac{2\pi e^2}{h}t^2 \lim_{\omega \to 0}
\frac{\mbox{Im} \chi_{ops}(\omega)}{\omega}, \label{eq:conductance}
\end{equation}
where $\chi_{ops}(\omega)$ is the analytic continuation
($i\omega_n \rightarrow \omega + i0$) of
the dynamical ``orbital pseudo-spin"  susceptibility for the
SU(4) Anderson Hamiltonian $H_A$ (without $H_T$),
\begin{eqnarray}
\chi_{ops}(i\omega_n) = \int_{0}^{\beta}d\tau e^{i\omega_n}
<{\cal T} \hat{T}_+(\tau)\hat{T}_-(0)>,
\end{eqnarray}
with the time-ordering operator ${\cal T}$.
The corresponding SU(2) operators are defined as
\begin{eqnarray}
\hat{T}_z & \equiv& \frac{1}{2}(\hat{n}_R-\hat{n}_L),\nonumber\\
\hat{T}_{\pm} & \equiv &\sum_{\sigma} |\sigma \pm1/2 \rangle \langle
\sigma \mp1/2 |.
\end{eqnarray}
These orbital pseudo-spin operators properly
describe inter-dot charge fluctuations.
As defined above, the eigenvalue $\tau=\pm 1/2$ of $T_z$
specifies which dot an electron occupies.
Eq.(\ref{eq:conductance}) means that the low-frequency inter-dot
``orbital" susceptibility is essential
to determine the conductance.

Although the low-frequency susceptibility is difficult to calculate in general,
we can make use of sophisticated techniques developed in the study of
the NMR relaxation rate in  dilute magnetic alloys \cite{shiba,nakamura}: 
the exact Ward-Takahashi relation for the low-frequency dynamical
pseudo-spin susceptibility is obtained, at zero temperature, as
\begin{eqnarray}
&&\lim_{\omega \to 0} \frac{\mbox{Im}\chi_{ops}
(\omega)}{\pi\omega} =
-\sum_{\sigma}\sum_{\tau, \tau'}(T_+)_{\tau\tau'}^2
K_{\tau \tau'}^\sigma,
\label{dynamical}
\end{eqnarray}
with
\begin{eqnarray}
K_{\tau \tau'}^\sigma=
\left\{
\begin{array}{rl}
\rho_{d\tau}^{\sigma} \, \rho_{d\tau'}^{\sigma}
\left[ 1+ \frac{\Sigma_{d \tau}^{\sigma}
- \Sigma_{d \tau'}^{\sigma}}{\varepsilon_{d \tau}^{\sigma}
- \varepsilon_{d \tau'}^{\sigma}} \right]^2  \, \,
(\mbox{for}\ \varepsilon_{d \tau}^{\sigma} \neq
\varepsilon_{d \tau'}^{\sigma}) \\
\left[  \chi_{ops}^\sigma (0)\right]^2  \,\,
(\mbox{for}\ \varepsilon_{d \tau}^{\sigma} =
\varepsilon_{d \tau'}^{\sigma})
\end{array} \right. \label{eq:korringa}
\end{eqnarray}
where
$\rho_{d\tau}^{\sigma}$ ($ \Sigma_{d\tau}^{\sigma}$)
is the density of states (self energy) for an electron
at the Fermi level in the dot $\tau$.

We note that the second line of Eq.(\ref{eq:korringa}) is
the well-kwon Korringa relation \cite{shiba}  in the context of
NMR relaxation theory, and the first line is its extension to the
case having a finite energy-level splitting \cite{nakamura}.
Since the static susceptibility can be calculated by the
exact solution, we need to evaluate the density of states
$\rho_{d\tau}^{\sigma}$ and the self energy $ \Sigma_{d\tau}^{\sigma}$.
Fortunately, this can be done by
exploiting the Friedel sum rule. First recall that
the phase shift $\delta_{\tau}^{\sigma}$ of an
electron in the double dots at the Fermi level  is
obtained from the
average number of electrons $\langle n_{d\tau}^{\sigma} \rangle$ in the
double dots: $\delta_{\tau}^{\sigma} = \pi \langle n_{d\tau}^{\sigma}\rangle$
(Friedel sum rule). Then, the density of states and the self energy
at the Fermi level is given by
\begin{eqnarray}
\rho_{d \tau}^{\sigma} &=& \frac{\sin^2 \delta_{\tau}^{\sigma}}{\pi \Delta} ,
\label{eq:fermil1} \\
\Sigma_{d \tau}^{\sigma} &=& \Delta \cot
\delta_{\tau}^{\sigma} - \varepsilon_{d \tau}^{\sigma} \label{eq:fermil2}
\end{eqnarray}
with the resonance width
$\Delta$ due to the mixing $V$.
Note that the electron number $\langle n_{d\tau}^{\sigma} \rangle$
can be evaluated by the exact solution.

Combining all the above relations, we can compute
the conductance at zero temperature:
the quantities in the right hand side of Eq.(\ref{eq:korringa}),
$\rho_{d\tau}^{\sigma}$, $\Sigma_{d \tau}^{\sigma}$,
$\chi^{\sigma}_{ops}(0)$, can be evaluated by means of
the Bethe-ansatz solution of the SU(4) Anderson
model \cite{schkawa}.

\section{double-dot system}\label{sc:doubl-dotresult}

\begin{figure}
\includegraphics[width=8.0cm]{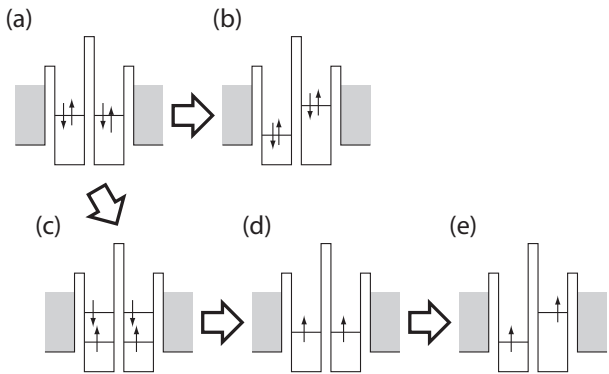}
\caption{Schematic description of the energy states in
the double dots connected in series: (a) SU(4) symmetric
case, (b) asymmetric
case, (c) symmetric case in a magnetic field, (d) symmetric
case in a strong magnetic field, (e) asymmetric
case in a strong magnetic field.}
\label{fig:energystates}
\end{figure}

We study the conductance in several cases in our double-dot
system, which
are schematically shown in FIG. \ref{fig:energystates}.

\subsection{Charge fluctuations in symmetric double dots}

\begin{figure}[bt]
\includegraphics[width=6.5cm]{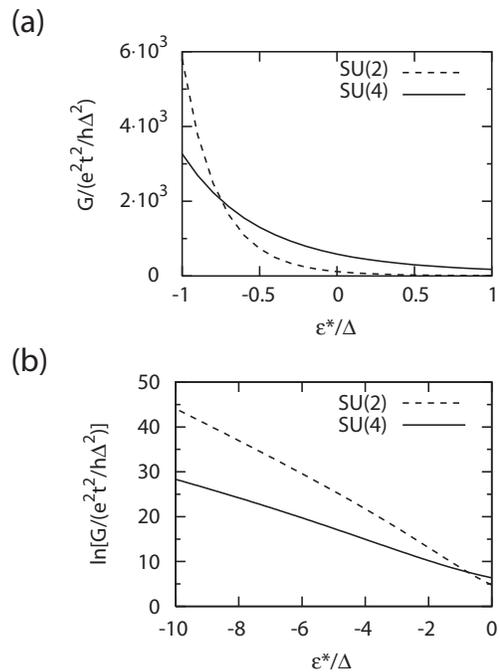}
\caption{(a)  Conductance in the cases of the SU(4) and SU(2) symmetric
double-dot systems as a function of the renormalized energy
level $\varepsilon^*$.
(b) conductance on log scale: we can see distinct exponential
dependence between the SU(4) (zero field) and SU(2)
(strong field) cases  in the Kondo regime.}
\label{fig:vf}
\end{figure}

Let us start with the double-dot system shown
in FIG. \ref{fig:energystates}(a), where the energy levels
of two dots are same, which we refer to as the
symmetric dots in this paper: there are
4 degenerate electron states including spin degrees of freedom.
In this case, from the
expressions (\ref{eq:conductance}), (\ref{dynamical}) and (\ref{eq:korringa})
we write down the conductance
in the absence of the magnetic field as,
\begin{eqnarray}
G &=& 2\pi^2 \cdot \frac{e^2}{h}t^2 \sum_{\sigma}
\left[ \chi^{\sigma }_{ops}(0) \right]^2 \nonumber \\
&=& 4\pi^2 \cdot \frac{e^2}{h}t^2
\left[ \chi^{\uparrow }_{ops}(0) \right]^2 .
\end{eqnarray}
By computing   static pseudo-spin susceptibility
$\chi^{\uparrow}_{ops}(0)$
by means of the Bethe-ansatz solution of the SU(4)
Anderson model, we evaluate
the conductance as a function of the effective
energy level $\varepsilon^*$ \cite{eflevel}.
The results are shown in FIG. \ref{fig:vf}.
When the dot-level $\varepsilon^*$ is above the Fermi level,
the conductance is small, since the resonant tunneling
does not occur. As  $\varepsilon^*$ goes down through the Fermi
level,  the conductance is enhanced by
the Kondo effect, which is analogous to an ordinary
single dot case.  However, in contrast to the single dot system,
for which the conductance is saturated in the Kondo limit with deep
$\varepsilon^*$,
it continues to increase exponentially.  The increase
is caused by the inter-dot charge fluctuations enhanced by the
inter-dot ``orbital"  Kondo effect
\cite{Wilhelm,Sun,Borda}. Since the
static susceptibility $\chi_{ops}^\sigma(0)$
is inversely proportional to the Kondo temperature
$T_K \sim \exp(-\Delta/\varepsilon^*)$,
the conductance has the exponential dependence like
$\exp(-2\Delta/\varepsilon^*)$.

Note that  the ordinary spin Kondo effect and the
inter-dot Kondo effect both emerge in the above SU(4) symmetric
case.  Therefore, in order to see the above
characteristic enhancement of the conductance more clearly,
we consider an extreme case with strong
magnetic fields, where the
spin Kondo effect is completely suppressed.
Shown in FIG. \ref{fig:vf}(b) is the conductance in strong
magnetic fields (corresponding to FIG. \ref{fig:energystates}(d)).
We can see the enhancement of the conductance due to
inter-dot Kondo effect with SU(2) symmetry. In this case, the corresponding
Kondo temperature is given by $T_K \sim \exp(-2\Delta/\varepsilon^*)$, so that
the increase of the conductance, $\sim \exp(4\Delta/\varepsilon^*)$,
is more significant in comparison
with the zero field case.
These results are  indeed seen in log-scale plots
given in FIG. \ref{fig:vf}, which clearly features
the exponential dependence
of the conductance in the Kondo regime.

\subsection{Symmetric double dots: magnetic-field control}

\begin{figure}[bthp]
\includegraphics[width=6.0cm]{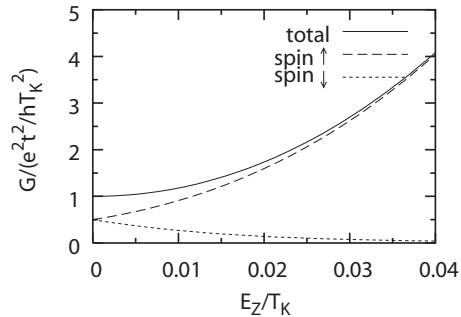}
\caption{Conductance as a function of the Zeeman splitting
in the Kondo regime.
We also show the contribution of the electrons with spins
parallel (anti-parallel) to the magnetic field.
Note that
$T_K=T_K(E_Z=0)$ is the Kondo
temperature of the  SU(4) Anderson model.}
\label{fig:ddmg}
\end{figure}

\begin{figure}[bthp]
\includegraphics[width=5.5cm]{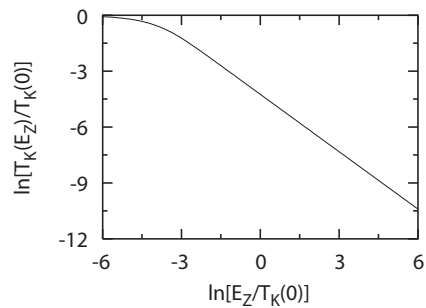}
\caption{The effective
Kondo temperature as a function of the Zeeman splitting on
log-log scale.}
\label{fig:loglogkt}
\end{figure}

It is seen from FIG. \ref{fig:vf} that in the Kondo regime with
deep dot levels, the conductance in  the SU(2) case
(strong field) is larger than that in the SU(4) case (zero field).
This implies that the conductance may be monotonically enhanced
in the presence of a magnetic field.  To clarify
this point, we focus on the field-dependence of the
conductance for the SU(4) symmetric double-dot system (shown
in FIG. \ref{fig:energystates}(c)) in the Kondo regime.
Following the way outlined above, we can derive the conductance
in this case,
\begin{eqnarray}
G = 2\pi^2 \cdot \frac{e^2}{h}t^2 \left[ \left\{ \chi^{\uparrow }_{ops}(0)
\right\}^2 + \left\{ \chi^{\downarrow }_{ops}(0) \right\}^2 \right].
\end{eqnarray}
By exploiting the exact solution of the SU(4) Anderson model in the
Kondo regime (so-called Coqblin-Schrieffer model), we compute
the conductance
as a function of the Zeeman splitting $E_Z$, which is shown
in FIG. \ref{fig:ddmg}. Also,
the effective Kondo temperature $T_K(E_Z)$ is plotted as
a function of the Zeeman splitting on the log-log scale
in FIG. \ref{fig:loglogkt}. Here we assume that the direction of the magnetic field is parallel to spin$\uparrow$. It is seen that the magnetic fields
enhance the conductance, in contrast to the ordinary
Kondo effect in a single-dot system.
The inter-dot Kondo effect is caused by the degenerate energy levels
in two dots, which still possess SU(2) symmetry in
strong magnetic fields.  Since the pseudo-spin susceptibility
$\chi_{ops}^{\uparrow}(0)$
increases with the increase of the field, thus resulting in the
enhanced conductance. In strong fields,
the effective Kondo temperature, which is defined by
the inverse of $\chi_{ops}^{\sigma}(0)$, is given by \cite{schkawa},
\begin{eqnarray}
T_K(E_Z)/T_K(0) \sim (E_Z/T_K(0))^{-1},
\end{eqnarray}
so that the conductance increases as
\begin{eqnarray}
G \sim (T_K(E_Z)/T_K(0))^{-2}\sim (E_Z/T_K(0))^{2}.
\end{eqnarray}
Here we note
that the conductance for electrons
with spin parallel (anti-parallel)
to the magnetic field increases (decreases).
This effect might be utilized for spin-current control
by using double-dot systems.
The above results agree with those of
Borda \textit{et al.} obtained  by the
numerical renormalization group analysis \cite{Borda}.

It is to be noted here that 
the SU(4) Kondo resonance has been observed not only in a
double-dot system \cite{Wilhelm} but also 
in a single vertical quantum dot whose symmetric shape
 gives rise to SU(4) internal degrees of freedom
\cite{Sasaki2}. Also, STM experiments on a Cr(001) surface
have found the SU(4) Kondo resonance, where the degeneracy 
of $d_{xz}$ and $d_{yz}$ states gives the additional orbital
degrees of freedom \cite{Zhuravlev}.
Our results are also consistent with these findings.

\subsection{Asymmetric double dots: gate-voltage control}

Next, we consider how the conductance is influenced by
the energy-level difference between
the two dots, which is controlled by changing the gate voltage
of each dot. We study two typical cases in the Kondo regime:
zero magnetic field  (FIG. \ref{fig:energystates}(b))
and strong magnetic fields (FIG. \ref{fig:energystates}(e)).

From the expressions (\ref{eq:conductance})
and (\ref{dynamical})-(\ref{eq:fermil2}), the conductance at zero field
is written as
\begin{eqnarray}
G=4 \cdot \frac{e^2}{h}t^2\frac{\sin^2(\pi\langle n_{L}^{\uparrow}
\rangle - \pi\langle n_{R}^{ \uparrow } \rangle)}{\delta E^2}.
\end{eqnarray}
We compute the conductance as a function
of energy differences $\delta E$.  We also study the conductance
in strong magnetic fields, where the system is completely polarized,
and the remaining inter-dot charge fluctuations are
described  by the SU(2) Kondo model subjected to the energy
difference $\delta E$\cite{Wilhelm,Sun,Borda}.
\begin{figure}[btp]
\includegraphics[width=6.0cm]{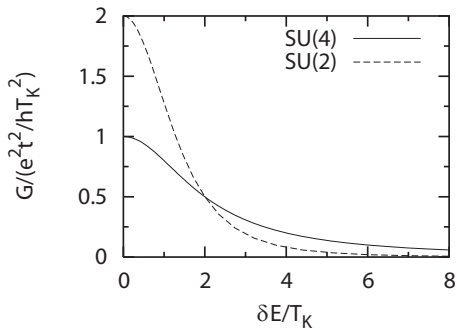}
\caption{Conductance as a function of the energy difference
$\delta E$ between two dots. We
take $T_K=T_K(E_Z=0)$ for the SU(4) double-dot case
(zero field)
and $T_K=T_K(E_Z=\infty)$ for the SU(2) double-dot case
(strong fields).}
\label{fig:dif}
\end{figure}
The results obtained in both cases are shown in FIG. \ref{fig:dif}.
In contrast to  the magnetic-field dependence discussed above,
the conductance decreases  monotonically as a function of
$\delta E$. The decrease is caused by the suppression of
the inter-dot Kondo effect in the presence
of finite energy difference $\delta E$.

We note here that the ordinary ``spin" Kondo effect
still persists even in finite $\delta E$, giving rise to the
enhanced spin fluctuations.  Although such enhancement in spin
fluctuations may not be
observed in transport properties, it should show up if we observe
the NMR relaxation rate in the double-dot system.  For example,
the $\delta E$-dependence of the NMR relaxation rate is
exactly given by the function shown in FIG. \ref{fig:ddmg}:
for large $\delta E$, it is enhanced as $(\delta E)^2$.

\section{triple-dot system}\label{sc:maltiple}

We now discuss how the above method can be used to calculate the
conductance for multiple-dot systems with more than two dots. Here,
we deal with a triple-dot system, and then briefly outline how to
extend the method to $N$-dot systems.
We will see that the conductance exhibits some characteristic
properties
under the control of the gate voltage and the magnetic field.

\begin{figure}[btp]
\includegraphics[width=4.0cm]{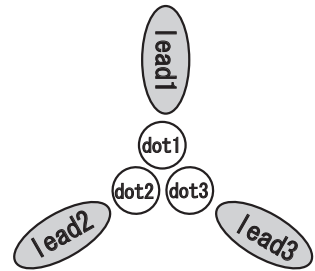}
\caption{Schematics of our triple-dot system: three dots are
connected via small tunneling $t$, and each dot is connected to a
lead via tunneling $V$. Inter- as well as intra-dot
Coulomb repulsions are assumed to be sufficiently strong.}
\label{fig:tripledot}
\end{figure}

Let us consider a triple-dot system, for which  three dots
and three leads as arranged as shown in FIG. \ref{fig:tripledot}.
Inter-dot tunneling $t$  (intra- and inter-dot Coulomb repulsions)
is assumed to be sufficiently small (large) here again. Therefore,
one of the three dots can accommodate an electron thanks to
strong intra- and inter-dot correlations.
We focus on the Kondo regime, where the energy levels
in the dots are sufficiently lower than the Fermi level.
We note that a similar but different triple-dot system has been
proposed recently \cite{kuzmenko} and
its symmetry properties have been discussed.

In the second order in tunneling $t$, we can  calculate
the conductance between two leads in the triple-dot system
by the exact solution of the SU(6) Anderson model, because
there are 6 available electron states including spin
degrees of freedom  in the three dots.
In this case, we can still utilize the formula Eq.(\ref{eq:conductance})
to calculate the conductance,
where the strong inter-dot correlations among three dots are
incorporated via the inter-dot susceptibility $\chi_{ops}$
between two dots through which electric currents flow.

\subsection{gate-voltage control}

\begin{figure}[btp]
\includegraphics[width=6.0cm]{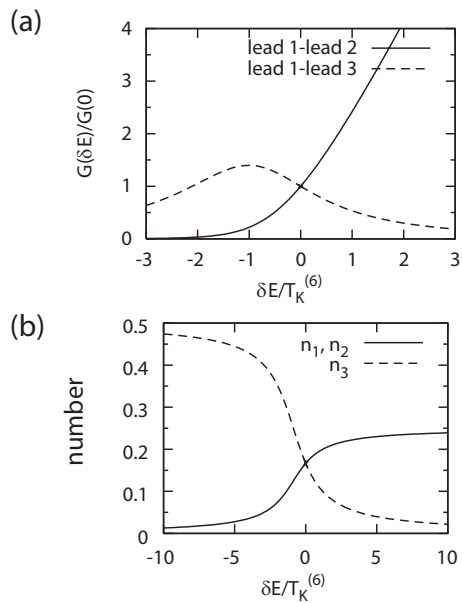}
\caption{(a)Conductance as a function of the  energy difference. Here, $T_K^{(6)}$ is the
Kondo temperature
for SU(6) triple-dot systems. (b)the number of electrons for the asymmetric
triple-dot  as a function of the energy difference. The solid line is
the number of electrons in the dot 1 or the dot 2 per spin and
the dashed line is that in the dot 3.}
\label{graph:tricon}
\end{figure}

Transport properties for the above three-dot systems
depend on how the current is observed.
To be specific, we  change the gate voltage
attached to the dot 3 with keeping the voltage in the
dots 1 and 2  fixed, and observe the conductance between
the leads 1 and 2 as well as between the leads 1 and 3.
\\
\\
The computed conductance
is shown in FIG. \ref{graph:tricon}(a) as a function of
the energy difference  $\delta E$  between the energy level in
the dot 3 and those in the dots 1 and 2.
We set the sign of $\delta E$ positive when
the energy level in the dot 3 is higher than
the others.

Let us first observe the current between the leads 1 and 2.
It is seen that the conductance increases
with the increase of $\delta E \, (>0)$.  This
increase is attributed to the enhancement of the inter-dot
Kondo effect in the presence of the energy deference, which is
similar to that for the
double dots in magnetic fields discussed in the previous section.
At $\delta E=0$, the current flows via an SU(6) Kondo resonance
(i.e. 6-fold degenerate Kondo resonance). On the other hand, for
large $\delta E$, the SU(4) Kondo effect is realized within
4 lower states in the dots 1 and 2. This gives the enhancement
of the inter-dot susceptibility between the dots 1 and 2,
resulting in the increase of the conductance. According to
the exact solution of the SU(6) Anderson model \cite{sch2},
the effective Kondo temperature
for large $\delta E$ is given by,
\begin{eqnarray}
T_K^{(eff)} (\delta E)/T_K^{(6)} \sim  (\delta E/T_K^{(6)})^{-1/2},
\end{eqnarray}
and the corresponding conductance is
\begin{eqnarray}
G \sim \delta E/T_K^{(6)},
\end{eqnarray}
where $T_K^{(6)}=T_K^{(eff)}(\delta E=0)$ is the Kondo temperature for SU(6) triple-dot systems.

On the other hand, for
$\delta E <0$, the SU(2) spin Kondo effect occurs in the lower
2 levels in the dot 3,
while the spin Kondo effects in the dots 1 and 2  are suppressed
because the number of electrons in the dots 1 and  2
decreases (see FIG. \ref{graph:tricon}(b)).
Also, the inter-dot Kondo effect
between  the dots 1 and  2 is suppressed. As a result
the conductance decreases when the current is observed
between leads 1 and 2.

If the current is observed between the leads 1 and 3, distinct
properties appear in the conductance.
As seen from  FIG. \ref{graph:tricon}(a),  for large $|\delta E|$
(irrespective of its sign),
the conductance decreases because the energy difference suppresses
the inter-dot Kondo effect  between the  dots 1 and 2. Notice that around
$\delta E/T_K^{(6)} \sim -1$, the conductance
has a maximum structure, where charge fluctuations between the
dots 1 and  3 are slightly enhanced. Anyway,  the
conductance exhibits behavior similar to that observed in
the double-dot case under the gate-voltage control.

\subsection{magnetic-field control}

\begin{figure}[btp]
\includegraphics[width=5.0cm]{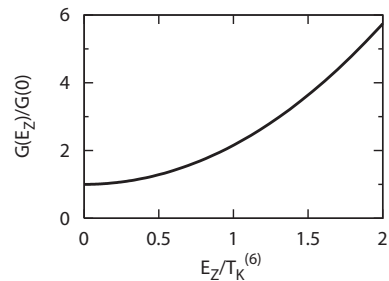}
\caption{Conductance as a function of the Zeeman splitting.}
\label{graph:triconmag}
\end{figure}

Let us now discuss how a magnetic field  affects transport properties.
For simplicity, we assume that the
energy levels of three dots are same (symmetric dots).
The computed conductance between two leads
under magnetic fields is shown
in FIG. \ref{graph:triconmag}.
The conductance increases
as the Zeeman splitting $E_Z$ increases although the Kondo effect due to spin fluctuations
are suppressed by the field.  As discussed in the
previous section, this enhancement is caused by
the inter-dot Kondo effect among three dots. For large magnetic fields,
half of the internal degrees of freedom are quenched, so that
the symmetry of the system changes from SU(6) to SU(3).
As a result, the SU(3) Kondo effect caused by inter-dot charge
fluctuations
is enhanced, and therefore the conductance is increased.
The effective Kondo temperature in large fields is given  as,
\begin{eqnarray}
T_K^{(eff)}(E_Z) /T_K^{(6)} \sim  (E_Z/T_K^{(6)})^{-1},
\end{eqnarray}
and thus the conductance is enhanced like
\begin{eqnarray}
G  \sim (E_Z/T_K^{(6)})^{2}.
\end{eqnarray}

\subsection{generalization to systems with more dots}

We can generalize our method to systems with more than three dots:
a lead is attached to each dot, where the electrons feel
strong intra- and inter-dot Coulomb repulsions.
All the dots are  connected to each other
via small inter-dot coupling $t$.

In similar manners mentioned above, we can calculate the conductance
in such multiple-dot systems. The calculation can be
done by using the formula Eq.(\ref{eq:conductance}) in the second
order in  tunneling $t$, where all the correlation effects are
incorporated through the dynamical susceptibility.
We can use the exact solution of the SU($2N$) Anderson model
\cite{schkawa}
for an $N$-dot system.
The conductance shows similar properties to those
observed  in the double and triple dots: if we change the gate voltage of the dot $\tau$, the conductance between the lead $\tau$ and one of the other leads is generally suppressed, while it is enhanced otherwise.

In this paper we have assumed small inter-dot
coupling $t$, and calculated the conductance up to
$t^2$. It should be mentioned that for a system with
more than two dots, a Fano-type interference effect may
emerge in higher order terms in $t$. This interference effect
may give another interesting aspect of multiple-dot
systems, which is to be studied in the future work.

\section{Summary}\label{summary}

We have studied transport properties in the double-dot system
connected in series that possesses not only intra- but also
inter-dot Coulomb repulsions. It has been shown that the
application of the Ward-Takahashi
identity enables us to use the exact solutions of the
Anderson model for calculations of the conductance at zero
temperature.  We have clarified
how the inter-dot Kondo effect affects the conductance
under the gate-voltage control  and the magnetic-field
control.  In particular, the conductance is
decreased by the suppression of the
inter-dot Kondo effect in the gate-voltage control, whereas
it is increased by the enhanced Kondo effect in
the presence of magnetic fields. The latter conclusion is consistent
with the results of the numerical renormalization
group. The method has also been applied to calculate
the conductance in multiple-dot systems including more
than two dots. By taking  a triple-dot
system as an example, we have shown how the conductance is controlled
by tuning the inter-dot Kondo effect.

Naively, it seems not easy to observe the Kondo effect
in multiple-dot systems  (more than two dots) experimentally.
We would like to mention, however, that the Kondo
temperature in multiple-dot systems
can be much higher than that in
single-dot systems when the inter-dot repulsion is relevant,
as assumed in this paper.  Therefore, if such multiple-dot
system could be fabricated, the Kondo effect may be possibly
observed even in multiple-dot systems considered here.

Finally a comment is in order on the ordinary ``spin" Kondo effect in
our multiple-dot system. We have focused on the inter-dot ``orbital"
Kondo effect in this paper, which directly affects transport properties.
Concerning the spin Kondo effect,
the impacts of the gate voltage
and the magnetic field appear differently from the
inter-dot  Kondo effect, e.g. the magnetic field
(gate-voltage difference) suppresses (enhances)
the spin Kondo effect. If we use the dynamical spin
susceptibility instead of the pseudo-spin susceptibility, the present
analysis can be straightforwardly applied to low-frequency
dynamics such as
the NMR relaxation rate, which may be important to discuss
an application to quantum-bits in quantum
computation. In fact, the expression Eq.(\ref{dynamical})
gives the NMR relaxation rate for the double-dot system,
if the dynamical susceptibility is regarded as the spin
susceptibility.  It is of particular interest that the NMR
relaxation rate in our multiple-dot systems can be controlled by
the gate voltage, e.g. the difference in the energy-levels
of double dots can enhance the relaxation rate.
\\
\\

After the completion of this paper, we became aware of a recent preprint which deals with the SU(4) Kondo effect in a slightly different model\cite{Chud}.

\begin{acknowledgments}
We would like to express our sincere thanks to M. Eto
for valuable discussions.
R.S. also thanks H. Akai for fruitful discussions and supports.
\end{acknowledgments}


\end{document}